\documentclass[seceq]{ptptex}

\newcommand\suzaku{{\it Suzaku}}

\newcommand\chandra{{\it Chandra}}

\newcommand\xmm{{\it XMM-Newton}}

\newcommand\s{{\rm~s}}
\newcommand\ks{{\rm~ks}}

\newcommand\kev{{\rm~keV}}

\newcommand\kms{\ifmmode {\rm~km\ s}^{-1} \else ~km s$^{-1}$\fi}
\newcommand\Hunit{\ifmmode {\rm~km\ s}^{-1}\ {\rm Mpc}^{-1}
        \else ~km s$^{-1}$ Mpc$^{-1}$\fi}
\newcommand\ctssec{\ifmmode {\rm~count\ s}^{-1} \else ~count s$^{-1}$\fi}
\newcommand\ergsec{\ifmmode {\rm~erg\ s}^{-1} \else
        ~erg s$^{-1}$\fi}
\newcommand\funit{\ifmmode {\rm~erg\ s}^{-1}\;{\rm cm}^{-2} \else
        ~ergs s$^{-1}$ cm$^{-2}$\fi}
\newcommand\phflux{\ifmmode {\rm~photon\ s}^{-1}\;{\rm cm}^{-2}
        \else   ~photon s$^{-1}$ cm$^{-2}$\fi}
\newcommand\efluxA{\ifmmode {\rm~erg\ s}^{-1}\;{\rm cm}^{-2}\;{\rm
        \AA}^{-1} \else ~erg s$^{-1}$ cm$^{-2}$ \AA$^{-1}$\fi}
\newcommand\efluxHz{\ifmmode {\rm~erg\ s}^{-1}\;{\rm cm}^{-2}\;{\rm
        Hz}^{-1} \else ~erg s$^{-1}$ cm$^{-2}$ Hz$^{-1}$\fi}
\newcommand\cc{\ifmmode {\rm~cm}^{-3} \else cm$^{-3}$\fi}
\newcommand\FWHM{\ifmmode {\rm~FWHM} \else ${\rm~FWHM}$\fi}
\newcommand\Msun{\ifmmode M_{\odot} \else $M_{\odot}$\fi}
\newcommand\Lsun{\ifmmode L_{\odot} \else $L_{\odot}$\fi}

\newcommand\hbeta{\ifmmode {\rm H}\beta \else H$\beta$\fi}
\newcommand\Kalpha{\ifmmode {\rm K}\alpha \else K$\alpha$\fi}
\newcommand\nh{\ifmmode N_{\rm H} \else N$_{\rm H}$\fi}
\usepackage{graphicx}
\usepackage{here}

\title{%        %You can use \\ for explicit line-break
Time lags in Narrow-line Seyfert 1 galaxies and the origin of their soft excess emission
}

\author{%       %Use \scshape  for the family name
G. C \textsc{Dewangan$^1$},
R. E. \textsc{Griffiths$^1$},
A. R. \textsc{Rao$^2$} \&
 \textsc{Surajit Dasgupta$^2$}
}

\inst{%     %Affiliation, neglected when [addenda] or [errata]
$^1$~Department of Physics, Carnegie Mellon University, 5000 Forbes Avenue, Pittsburgh, PA, 15213 USA \\
$^2$~Department of Astronomy \& Astrophysics, Tata Institute of Fundamental Research, Homi Bhabha Road, Mumbai 400005 India
}

\abst{%         %this abstract is neglected when [addenda] or [errata]
The origin of soft X-ray excess emission from type 1 active galactic nuclei has remained a major problem for the last two decades. It has not been possible to distinguish alternative models for the soft excess emission despite the excellent data quality provided by \xmm{} and \chandra{}. Here we present observations of time lags between the soft and hard band X-ray emission and discuss the implications to the models for the soft excess. We also device a method to distinguish the models for the soft excess using \suzaku{}'s broadband capability. 
}

\begin{document}

\maketitle

\section{Introduction}
Many type 1 active galactic nuclei (AGN) show soft excess emission below $\sim 2\kev$ and above a hard power law. Narrow-line Seyfert 1 galaxies (NLS1s) show the extreme soft excess emission among all AGN. The soft excess emission is a smooth continuum component. It is not a blend of narrow emission/absorption features as revealed by \chandra{} and \xmm{} high resolution grating observations \cite{Turneretal01}. When described as a thermal component, the soft excess emission has nearly uniform temperature in the range of $\sim 0.1 - 0.2\kev$ for AGN across a wide range in their black hole masses and luminosities\cite{Czernyetal03,GD04,Crummyetal06}. The temperature of the soft excess emission is too hot to be optically thick emission from standard accretion disks around super-massive black holes.  The origin of soft excess emission has remained a major problem in AGN research for the last two decades.

 There are three possible models for the origin of the soft excess emission.  One possibility is that the soft excess can be produced by thermal Comptonization in a cool ($kT\sim 0.1\kev$) and optically thick ($\tau \sim 10$) corona. This medium is different from the hot corona responsible for the primary power-law continuum. The cool Comptonizing region could be the ionized skin of an accretion disk or an optically thick region between the disk and hot inner flow. Gierli{\'n}ski \& Done\cite{GD04} have examined this model and found inappropriate as the temperature of the cool Comptonizing region is nearly constant for AGN with a wide range in their black hole masses, luminosities and accretion rates. In 2002, Fabian et al.\cite{Fabianetal02} showed that Compton reflection from a partially ionized skin on the accretion disk can reproduce the observed soft excess emission. In this model, the soft excess emission is a blend of a large number of relativistically broadened  emission lines from the disk and enhanced reflection continuum due to partially ionized skin of the disk. In 2004, Gierli{'n}ski \& Done\cite{GD04}  suggested that the partially ionized material is seen in absorption rather than in reflection as in the previous model. Again a large velocity smearing is required to hide the atomic line and edge features. High velocity winds arising from the accretion disk can provide the required smearing.

The three models, discussed above, describe the observed data statistically equally well. The $\chi^2$ statistics cannot distinguish the models\cite{SD05}. Also the ionized reflection and smeared wind absorption models describe the {\it rms} variability spectrum quite well and it has not been possible to identify the best model\cite{Pontietal06,GD06}.     
Here we present cross-correlation between soft and hard band X-ray emission from the NLS1 galaxies Akn~564 and Mrk~110  and discuss implications of the measured time lags on the origin of the soft excess emission. We also discuss how \suzaku{} can be used to constrain the models for the soft excess emission.

\begin{figure}
 \centering
\includegraphics[width=3.5cm]{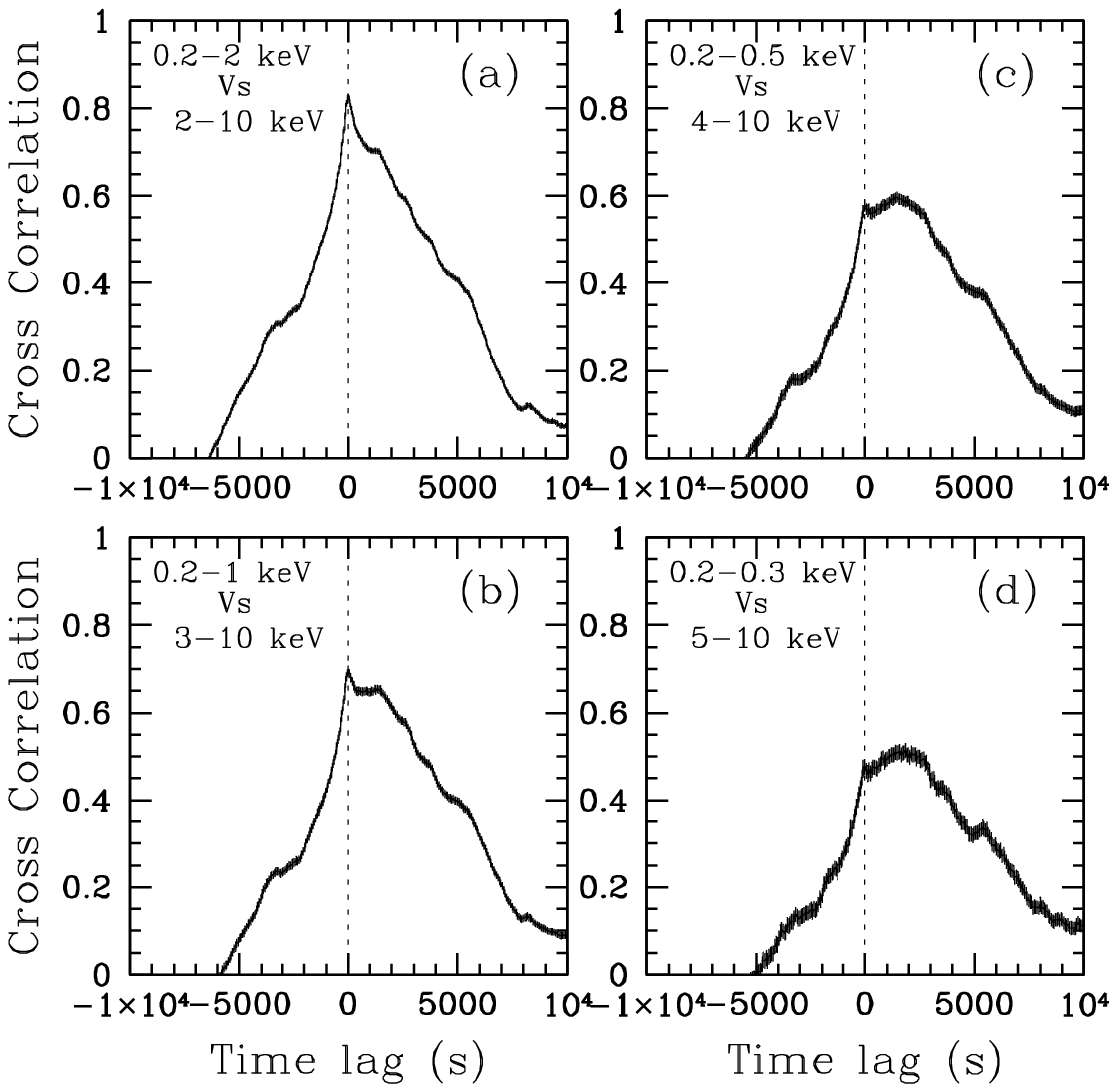}
\includegraphics[width=5cm]{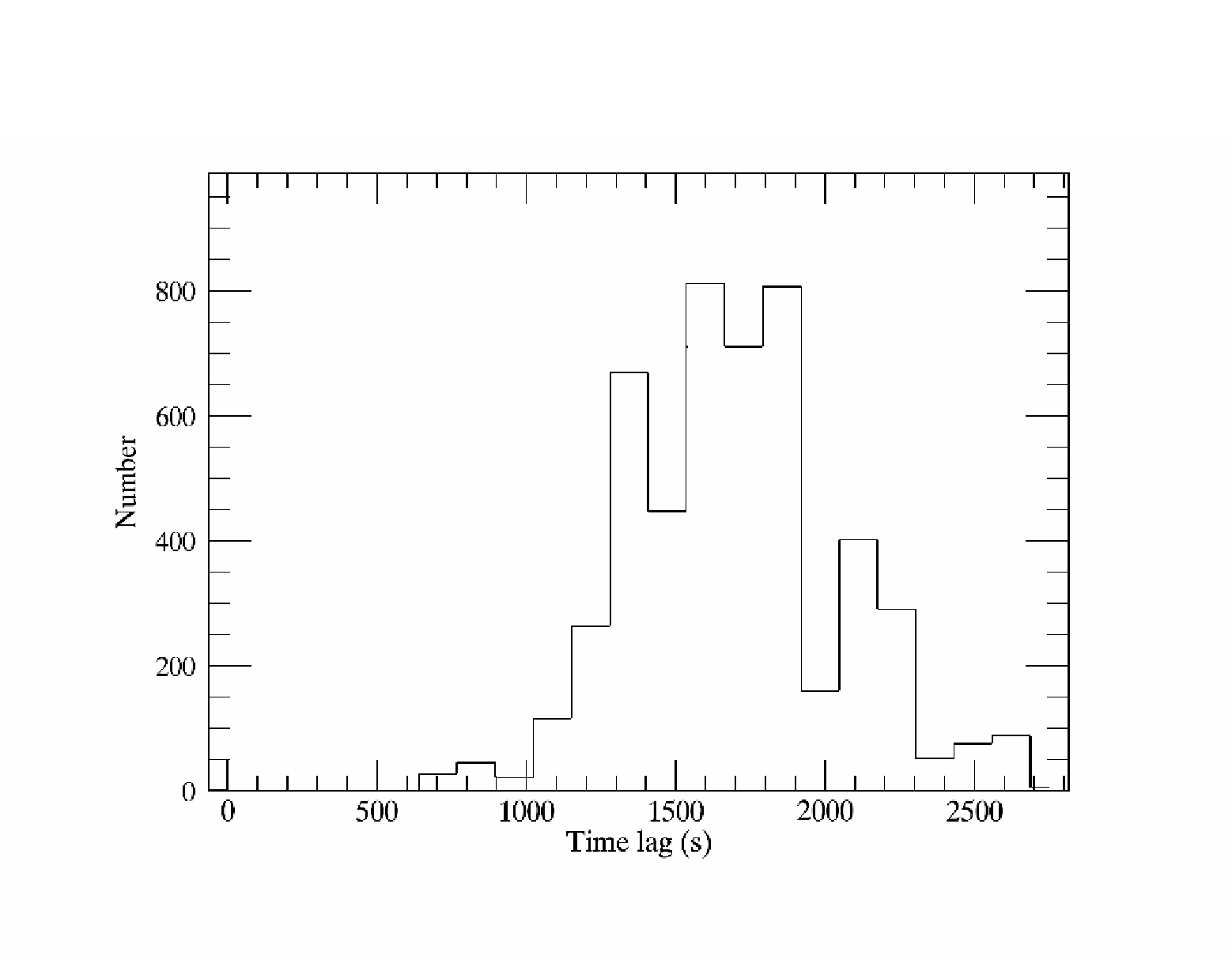}
  \caption{{\it Left:} Cross correlation function between soft and hard X-ray light curves of Akn~564. {\it Right:} Distribution of time lags between the $0.2-0.3$ and $5-10\kev$ bands derived from 5000 pairs of simulated light curves based on the observed light curves of Akn~564\cite{Dewanganetal07}.
}
  \label{f1}
\end{figure}
\begin{figure}
 \centering
\includegraphics[width=3.5cm,angle=-90]{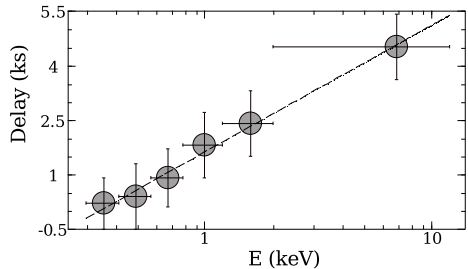}
\caption{Mrk~110: Time lag as a function of energy (reproduced from Dasgupta \& Rao (2006)\cite{DR06}}
\label{f2}
\end{figure}

\section{Time delays}
 Figure~\ref{f1}({\it left}) shows the cross-correlation function (CCF) between soft and hard band light curves, obtained from a $\sim100\ks$ \xmm{} observation, as a function of time delay. Positive delay means hard band lags the soft band. As the separation between the two bands increases, the correlation at zero lag disappears and a new correlation at positive delay ($\sim 2000\s$) appears.  It  is not easy to measure the time delay by simply fitting a function to the CCF. First, the shape of the CCF is complex and the errors on the CCF are unlikely to be normally distributed. To measure the time delay, we simulated 5000 pairs of soft and hard band light curves based on the error and count rate of the observed light curves and cross-correlated each pair. Figure~\ref{f1} ({\it right}) shows the distribution of the CCF peaks. We integrated this distribution to obtain $68\%$, $90\%$ and $99\%$ confidence ranges of the time lag, $\tau = 1728\pm336$, $1728\pm576$ and $1728\pm960\s$, respectively. There is also evidence for energy dependent time delay from Akn~564\cite{Arevaloetal06}. The lag between the $0.7-2\kev$ and $5-10\kev$ bands is about a factor of two larger than the lag between the $0.7-2$ and $2-5\kev$ bands. The lag spectrum of Akn~564 is very similar to that of Galactic black hole binaries (BHBs) in their very high states except that the time lag is about $\sim4-5$ orders of magnitude lower in BHBs\cite{Arevaloetal06}. 

Mrk~110 is another NLS1 galaxy that shows evidence for time delay\cite{DR06}. The time delay increases as the hard band  becomes harder. Figure~\ref{f2} shows the time lag as a function of energy. The delay is proportional to the $logE$, consistent with simple Comptonization model\cite{DR06}.

\section{Implications of the observed time delay} 
The observed hard band delays from Akn~564 and Mrk~110  demonstrate that the soft excess emission cannot be the reprocessed emission of hard X-ray emission.  
In the ionized reflection model, the soft excess emission is mainly a blend of a large number of broad emission lines from the disk i. e., the reprocessed emission. Most of the broad emission lines constituting the soft excess emission are excited by photons in the $1-2\kev$ band. Therefore, the soft excess emission below $0.5\kev$ cannot lead the $1-2\kev$ band. This is inconsistent with the observed time lag proportional to $logE$ in Mrk~110. Also the ionized reflection model is not a statistically acceptable fit to the X-ray spectrum of Akn~564\cite{Dewanganetal07}.
In the smeared wind absorption model, the soft excess is an artifact of the deficit of emission caused by the smeared wind absorption\cite{GD04,GD06}. This means the soft and hard band emission arise from the same continuum component or physical process - Comptonization. The time delay is a natural outcome of the Comptonization process as the soft photons undergo fewer scattering events than the hard photons. However, there three problems with the smeared wind model. ($i$) This model (XSPEC local model {\tt swind1}\cite{GD04}) does not describe the X-ray spectrum of Akn~564 satisfactorily\cite{Dewanganetal07}.  ($ii$) The $3-10\kev$ band shows more variability power on short time scales than the $0.2-0.5\kev$ band, which is not expected in the simple Comptonization models. ($iii$) Hydrodynamical simulations show that the line driven winds do not have the required smearing velocity (see the poster by Schurch \& Done). It remains to be seen if the magnetically driven winds can provide  the required smearing. 

\begin{figure}
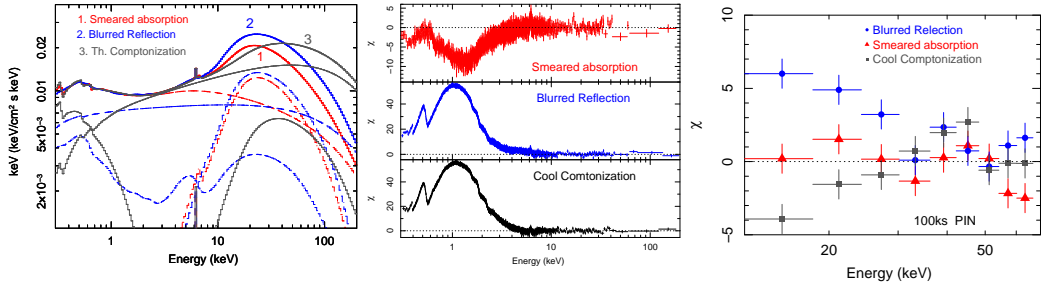

  \centering \includegraphics[width=3.5cm,angle=-90]{f3left.ps}
\includegraphics[width=3.5cm,angle=-90]{f3middle.ps}
\includegraphics[width=3.7cm,angle=-90]{f3right.ps}
\caption{{\it Left:} A comparison of the best-fit models, derived from the \xmm{} data, extended to high energies. {\it Middle:} Deviations of simulated data for \suzaku{} from the best-fit power-law model. {\it Right:} Deviations of the simulated data from the smeared wind absorption model. }
\label{f3}
\end{figure}

\section{Testing the models for the soft excess with \suzaku{}}
We have fitted the \xmm{} spectrum of Mrk~110 with the smeared wind absorption, blurred ionized reflection and thermal Comptonization models. The power law and Compton reflection from distant material were common to these models. The three models describe the observed data statistically equally well in the $0.3-10\kev$ band but differ significantly above $10\kev$.  The soft deficit in the smeared wind model requires steepest continuum while the ionized reflection model predicts the strongest Compton reflection emission. Figure~\ref{f3} shows the comparison of the best-fitting models, extrapolated to high energies. The three models predict distinct hard X-ray spectra in the \suzaku{} HXD PIN band.  We have simulated $200\ks$  \suzaku{} XIS+PIN spectra of Mrk~110 based on the three models derived from the \xmm{} data. Fig.~\ref{f3} ({\it middle}) shows the deviations of the simulated data from the best-fitting power law above $10\kev$. Clearly the smeared absorption model is distinguished from the other two models. The three models predict different amount of reflection as seen in the deviations of the simulated data from the smeared wind model in the HXD PIN band. Thus the \suzaku{}'s broadband capability can easily distinguish the three models for the origin of the soft excess emission.   

\section*{Acknowledgments}
It is pleasure to thank the LOC and SOC for organizing such a fruitful conference. GCD acknowledges the support of a NASA grant through the award NNX06AE38G. Most of the results presented here are based on observation obtained with \xmm{}, an ESA science mission with instruments and contributions directly funded by ESA Member states and NASA.

%\appendix
%\section{First Appendix} %Empty argument \section{} yields `Appendix'. 
%
%\section{Second Appendix}

\end{document}